\documentclass[
superscriptaddress,
amsmath,amssymb,
aps,pra,twocolumn,
]{revtex4-2}
\usepackage{graphicx}
\usepackage{dcolumn}
\usepackage{bm}
\usepackage{xcolor}
\usepackage{hyperref}
\usepackage[T1]{fontenc}
\hypersetup{colorlinks=true, citecolor=blue, urlcolor=blue, linkcolor=blue}
\newcommand{\ket}[1]{\left|#1\right\rangle}

\begin{document}

\title{Efficient photo-ionizing elimination of detrimental electric fields for Rydberg atoms}

\author{Zhou-Chen Deng}
\thanks{These authors contributed equally to this work.}
\affiliation{Hefei National Research Center for Physical Sciences at the Microscale and Department of Modern Physics, University of Science and Technology of China, Hefei 230026, China}
\affiliation{Shanghai Research Center for Quantum Science and CAS Center for Excellence in Quantum Information and Quantum Physics, University of Science and Technology of China, Shanghai 201315, China}

\author{Hao-Nan Lin}
\thanks{These authors contributed equally to this work.}
\affiliation{Hefei National Research Center for Physical Sciences at the Microscale and Department of Modern Physics, University of Science and Technology of China, Hefei 230026, China}
\affiliation{Shanghai Research Center for Quantum Science and CAS Center for Excellence in Quantum Information and Quantum Physics, University of Science and Technology of China, Shanghai 201315, China}

\author{Yu-Cheng Duan}
\affiliation{Hefei National Research Center for Physical Sciences at the Microscale and Department of Modern Physics, University of Science and Technology of China, Hefei 230026, China}
\affiliation{Shanghai Research Center for Quantum Science and CAS Center for Excellence in Quantum Information and Quantum Physics, University of Science and Technology of China, Shanghai 201315, China}

\author{Qi Zhang}
\affiliation{Hefei National Research Center for Physical Sciences at the Microscale and Department of Modern Physics, University of Science and Technology of China, Hefei 230026, China}
\affiliation{Shanghai Research Center for Quantum Science and CAS Center for Excellence in Quantum Information and Quantum Physics, University of Science and Technology of China, Shanghai 201315, China}

\author{Xiang-Can Cheng}
\affiliation{Hefei National Research Center for Physical Sciences at the Microscale and Department of Modern Physics, University of Science and Technology of China, Hefei 230026, China}
\affiliation{Shanghai Research Center for Quantum Science and CAS Center for Excellence in Quantum Information and Quantum Physics, University of Science and Technology of China, Shanghai 201315, China}
\affiliation{Hefei National Laboratory, University of Science and Technology of China, Hefei 230088, China}

\author{Yang Liu}
\affiliation{Hefei National Research Center for Physical Sciences at the Microscale and Department of Modern Physics, University of Science and Technology of China, Hefei 230026, China}
\affiliation{Shanghai Research Center for Quantum Science and CAS Center for Excellence in Quantum Information and Quantum Physics, University of Science and Technology of China, Shanghai 201315, China}

\author{Zhao-Yang Yuan}
\affiliation{Hefei National Research Center for Physical Sciences at the Microscale and Department of Modern Physics, University of Science and Technology of China, Hefei 230026, China}
\affiliation{Shanghai Research Center for Quantum Science and CAS Center for Excellence in Quantum Information and Quantum Physics, University of Science and Technology of China, Shanghai 201315, China}

\author{Jie Li}
\affiliation{Hefei National Research Center for Physical Sciences at the Microscale and Department of Modern Physics, University of Science and Technology of China, Hefei 230026, China}
\affiliation{Shanghai Research Center for Quantum Science and CAS Center for Excellence in Quantum Information and Quantum Physics, University of Science and Technology of China, Shanghai 201315, China}
\affiliation{Hefei National Laboratory, University of Science and Technology of China, Hefei 230088, China}

\author{Peng Liu}
\affiliation{Hefei National Research Center for Physical Sciences at the Microscale and Department of Modern Physics, University of Science and Technology of China, Hefei 230026, China}
\affiliation{Shanghai Research Center for Quantum Science and CAS Center for Excellence in Quantum Information and Quantum Physics, University of Science and Technology of China, Shanghai 201315, China}
\affiliation{Hefei National Laboratory, University of Science and Technology of China, Hefei 230088, China}

\author{Zhan Wu}
\affiliation{Hefei National Research Center for Physical Sciences at the Microscale and Department of Modern Physics, University of Science and Technology of China, Hefei 230026, China}
\affiliation{Shanghai Research Center for Quantum Science and CAS Center for Excellence in Quantum Information and Quantum Physics, University of Science and Technology of China, Shanghai 201315, China}
\affiliation{Hefei National Laboratory, University of Science and Technology of China, Hefei 230088, China}

\author{Chao-Yang Lu}
\affiliation{Hefei National Research Center for Physical Sciences at the Microscale and Department of Modern Physics, University of Science and Technology of China, Hefei 230026, China}
\affiliation{Shanghai Research Center for Quantum Science and CAS Center for Excellence in Quantum Information and Quantum Physics, University of Science and Technology of China, Shanghai 201315, China}
\affiliation{Hefei National Laboratory, University of Science and Technology of China, Hefei 230088, China}

\author{Jun Rui}
\affiliation{Hefei National Research Center for Physical Sciences at the Microscale and Department of Modern Physics, University of Science and Technology of China, Hefei 230026, China}
\affiliation{Shanghai Research Center for Quantum Science and CAS Center for Excellence in Quantum Information and Quantum Physics, University of Science and Technology of China, Shanghai 201315, China}
\affiliation{Hefei National Laboratory, University of Science and Technology of China, Hefei 230088, China}

\author{Jian-Wei Pan}
\affiliation{Hefei National Research Center for Physical Sciences at the Microscale and Department of Modern Physics, University of Science and Technology of China, Hefei 230026, China}
\affiliation{Shanghai Research Center for Quantum Science and CAS Center for Excellence in Quantum Information and Quantum Physics, University of Science and Technology of China, Shanghai 201315, China}
\affiliation{Hefei National Laboratory, University of Science and Technology of China, Hefei 230088, China}

\date{\today}

\begin{abstract}
Rydberg atoms are highly sensitive to external electric fields due to their exaggerated electronic properties. This unique feature underpins many of their applications in quantum science. However, uncontrolled stray electric fields can be detrimental, severely degrading quantum control. In this work, we demonstrate a novel scheme that efficiently creates an in-vacuum charge source by photo-ionizing laser-cooled atoms and eliminates detrimental electric fields in a Rydberg-atom tweezer array platform, using only readily available resources. With this method, we began with Stark-ionized Rydberg continuum spectra caused by a large, unknown stray electric field of more than $10~\mathrm{V/cm}$ and ultimately recovered well-resolved coherent excitation of individual Rydberg states after eliminating the residual field to an upper limit of $20(1)~\mathrm{mV/cm}$. Our method applies directly to existing Rydberg-atom platforms and other experiments sensitive to stray electric fields.
\end{abstract}

\maketitle


\section{Introduction}

Rydberg atoms, with electrons in highly excited states, exhibit extraordinarily large electronic wavefunctions and electric dipole moments between neighboring dipole-allowed states \cite{Gallagher2009}. This unique property leads to the extreme sensitivity of a single Rydberg atom's energy level to external DC or microwave electric fields \cite{Sedlacek_micrometer_2012, Jing_micrometer_2020, Zhang_electrometry_2026}, and more importantly, to the emergence of strong, long-range interactions between them \cite{Sibalic_Rydberg_2018}. In recent years, significant progress has been made in the quantum control of Rydberg atoms \cite{Levine_highfidelity_2018, Wilson_trap-ryd_2022, Muni_circular_2022, Anand_dual-species_2024}, making them a powerful platform for quantum science \cite{Saffman_RMP_2010}, including quantum nonlinear optics \cite{Liang_three-photon_2018, Lampen_super-atom_2019}, quantum metrology \cite{Finkelstein_clock_2024, Cao_clock_2024}, quantum simulation \cite{Browaeys_review_2020, Chen_AB_2025}, and computation \cite{Bluvstein_logical_2024}. Many achievements depend directly on the lifetime, coherence, and uniform excitation of Rydberg states throughout the system, making them extremely vulnerable to external perturbations \cite{Sylvain_ryd-error_2018}, particularly stray electric fields \cite{Hattermann_surface_2012, Hankin_efield_2014, Jia_Cavity_2016, Sheng_cavity_2017, Mamat_E-field_2024, Ocola_surface_2024}. In experiments, controlled electric fields have long been used for Stark ionization of Rydberg states \cite{Ducas_ionization_1975, Jeys_ionization_1980, Millen_Sr-ryd-spec_2011} or to precisely tune the Stark shift to engineer F\"{o}rster resonance between different Rydberg states \cite{Ravets_forster_2014, Anand_dual-species_2024}. However, an uncontrolled, drifting electric field can cause catastrophic effects, such as undesired energy shifts, dephasing, and strong mixing of Rydberg states, inevitably restricting or even preventing the use of Rydberg atoms as valuable quantum resources.

Such undesired stray electric fields may arise from photoelectric effects \cite{Rousseau_laser-charge_1968, Gaudart_Sr-film_1977, Duspayev_cell-voltage_2024, Patrick_vapor_2025} or adsorbate dipoles on nearby surfaces \cite{Safavi-Naini_chargemodel_2011, Sedlacek_adsorbate_2016} that are particularly acute in hybrid on-chip \cite{Tauschinsky_surface_2010, Carter_chip_2012, Hattermann_surface_2012, Ocola_surface_2024} or optical cavity systems \cite{Jia_Cavity_2016, Sheng_cavity_2017}. Several strategies have been developed to address the electric-field issue, but each has its own limitations. For example, the small number of electrons from black-body ionization of Rydberg atoms can cancel the electric field from surface Rb adsorbates \cite{Sedlacek_adsorbate_2016}, but the blockade effect in Rydberg excitation limits the charge flux. Ultraviolet (UV) illumination used for light-induced atomic desorption (LIAD) has also been applied to reduce adsorbate-related fields \cite{Davtyan_adsorbate_2018, Mamat_E-field_2024, Ziemba_UV-desorp_2025}, but its effectiveness depends delicately on the surface material and adsorbate species \cite{Chan_adsorbate_2014}. Photo-ionization of residual vacuum-background gas by vacuum-UV or soft-X-ray radiation can actively neutralize static charges in industrial high-vacuum settings \cite{Inaba_VUV_1994, Kim_VUV_2017}. However, the gas density is extremely low in typical ultrahigh-vacuum (UHV) Rydberg-atom apparatus, and dedicated optical coatings or glass materials do not efficiently transmit high-energy photons. Dedicated in-vacuum electrodes can partially compensate or shield the field, but they must be incorporated into the vacuum apparatus from the beginning and require frequent calibration \cite{Sylvain_ryd-error_2018, Anand_dual-species_2024, Panja_electrode_2024}. Therefore, developing more efficient, robust methods to address drifting stray electric fields is vital for achieving high-fidelity control over Rydberg states.

\begin{figure*}
\centering 
\includegraphics[width=0.95\textwidth]{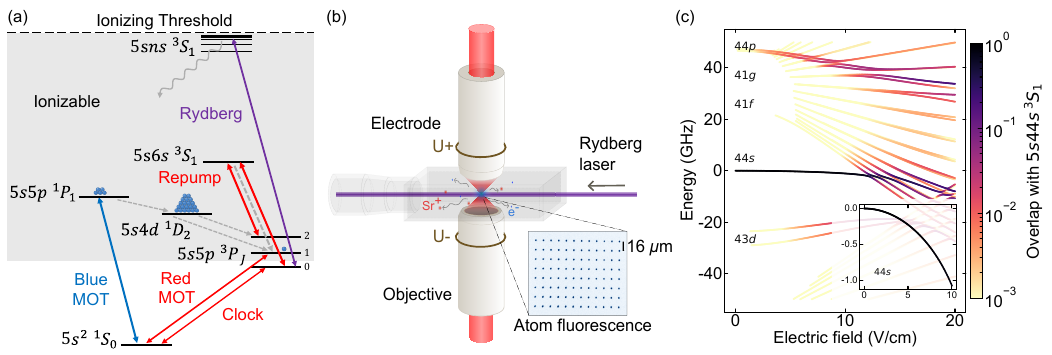}
\caption{\label{fig_setup_scheme}
Illustration of the MOT photo-ionization method for eliminating stray electric fields. (a) Relevant energy levels of $^{88}\mathrm{Sr}$ atoms. The Rydberg laser can directly photo-ionize atoms in excited states within the shaded region, which are continuously populated during the blue-MOT loading. The blue spheres indicate the approximate steady-state population ratio. (b) Experimental layout. A Rydberg laser from the side can photo-ionize atoms in a MOT inside a fused-silica glass cell with external dimensions of $35\times25\times130$ mm and a thickness of 3.5 (4.0) mm for the horizontal (vertical) surfaces. Only four rectangular outer surfaces are dielectric-coated. In addition to the stray electric field, we can generate a controlled field by applying static voltages across external copper loop wires wound around the objective to benchmark the field-reduction performance. (Inset) Average fluorescence image of single atoms in an $8\times12$ array. (c) Stark map of the $5s44s\,{}^3S_1$ Rydberg state calculated with PairInteraction~\cite{Weber_Pair-Interaction_2017}. We use $ns/nd$ throughout as shorthand for the outer-electron orbital only. An electric field can strongly shift the target state's energy and mix it with neighboring levels. We include basis states with $39\leq n\leq49$, $0\leq l\leq20$, and $-5\leq m_j\leq5$ in this calculation. }
\end{figure*}

In this work, we started with an unexpectedly large stray electric field that essentially Stark-ionized the highly excited Rydberg states, and conventional UV-light desorption \cite{Mamat_E-field_2024} was rather ineffective. Instead, we found that illumination with a UV light field can efficiently photo-ionize laser-cooled atoms trapped in a magneto-optical trap (MOT) \cite{Jones_photoassociation_RMP_2006, Henkel_Rb-ionize_2010}. This creates a large flux of in-vacuum charge sources that can efficiently eliminate the large stray electric field arising from the combined effects of surface adsorbate dipoles \cite{Sedlacek_adsorbate_2016}, photoelectric effects \cite{Gaudart_Sr-film_1977, Patrick_vapor_2025}, and static charges on optics outside the vacuum. The MOT photo-ionization method has an advantage in continuously replenishing laser-cooled atoms, compared to conventional approaches that rely on background gases or a small number of Rydberg atoms. This robust method does not depend on the microscopic mechanism of the electric field and can eliminate both stray surface-related and externally generated electric fields. After dynamical elimination, we achieved uniform, stable Rydberg-state excitation and estimated the remaining field with an upper limit of $20(1)~\mathrm{mV/cm}$ in its standard deviation. The scheme uses only readily available resources and applies directly to a broad range of existing Rydberg-atom experiments with similar issues.

\section{Rydberg atom array and the electric-field issue}

Our experiment begins with loading cold $^{88}\mathrm{Sr}$ atoms in a three-dimensional blue-MOT operating on the broad $5s^2\,{}^1S_0 \leftrightarrow  5s5p\,{}^1P_1$ transition at 461 nm, with transversely cooled atom fluxes generated by a custom-designed atom source \cite{Li_Sr-source_2023}. As shown in Fig. \ref{fig_setup_scheme}(a), in the MOT cooling cycle, a weak leakage channel populates the intermediate $5s4d\,^{1}D_2$ state, which further slowly decays to the $5s5p\, ^{3}P_J$ states \cite{Pucher_Sr_data_2025}. We used two repumping lights at 679 nm and 707 nm to quickly pump the atoms into the slowly decaying $^{3}\mathrm{P}_1$ state, which then returns to the ground state. With a rate-equation model (see Appendix \ref{app:rate-equation-model}), we estimate the populations in the $^{1}P_1$, ${}^1D_2$, and $^{3}P_1$ states have a ratio of about 5:20:1, enabling efficient single-photon ionization under a UV light field \cite{Haq_Sr-ionize_2006, Haq_Sr-ionize_2007}. We then transferred the atoms to a narrow-line red-MOT on the intercombination transition at 689 nm \cite{Pucher_Sr_data_2025}. The red-MOT further reduces the atomic temperature to $\sim 2\,\mu\mathrm{K}$ and prepares a total atom number of $\ sim 2\times10^ {5}$. The atoms were then stochastically loaded into an optical tweezer array operating at 813 nm, which was created using a spatial light modulator (SLM) \cite{Lin_AI-SLM_2025} through a high numerical aperture (NA) objective with NA = 0.55, as shown in Fig. \ref{fig_setup_scheme}(b). We used a 461 nm probe beam to induce light-assisted collisions for parity projection or to scatter fluorescence photons for imaging, while cooling the atoms with a single 689 nm Sisyphus cooling beam \cite{Covey_Sr813tweezer_2019}. At this point, we prepared an $8\times 12$ array with about 50$\%$ of the sites randomly filled with single atoms at a mean temperature of $3.6\,\mu$K. To transfer the atoms into the metastable clock state, we applied a uniform magnetic field of 350 G to induce the coherent $5s^2\,{}^1S_0 \leftrightarrow  5s5p\,{}^3P_0$ clock transition at 698 nm, with residual atoms optically pumped into the clock state. We then ramped the magnetic field down to about 100 G and directly excited the trapped atoms into highly excited Rydberg states using a frequency-locked, low-phase-noise 317 nm laser with a beam waist of about 100 $\mu$m. Throughout this work, we probed the spectroscopy with a randomly filled atom array with a large spacing to avoid Rydberg interaction effects. 

\begin{figure}[t]
\centering 
\includegraphics[width=\columnwidth]{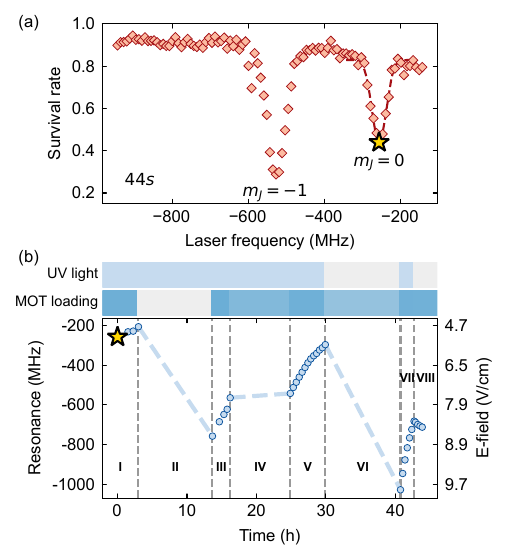}
\caption{\label{fig_resolve_ryd_drift}
Monitoring the drifts of the $5s44s\,{}^3S_1$ resonance frequency under a stray electric field. 
(a) First clearly resolved resonances of the target transition relative to the Rydberg $m_J=0$ level at zero electric field. The splitting between the resonances matched the Zeeman splitting of the $^3S_1$ Rydberg state under a magnetic field of 100 G.  
(b) Drifts in the resonance frequency of the $m_J=0$ state under different conditions. Each data point is fitted from a single spectroscopic scan; for time intervals without data points, we performed no spectroscopic scan. The grayness of the UV LED light or MOT loading indicates whether the light is on or off, or the sequence's duty cycle.  We observe that when MOT loading and UV LED light illumination are applied together, the resonance frequency increases significantly, indicating a reduced stray electric field. 
Error bars represent the standard error of the mean (s.e.m.) for the survival rate (a) and the standard deviation (std.) of the fitted resonance frequency (b), and both are smaller than the symbol sizes. }
\end{figure}

Unfortunately, in our first attempt to locate the transition of the $5s61s\,{}^3S_1$ Rydberg states, we could not resolve any resonances over a broad frequency range near the theoretically predicted frequencies \cite{madjarov_thesis_2021}. Our measurement shows that the Rydberg laser directly ionizes atoms in the $^3P_0$ state over a broad frequency range (see details in Appendix \ref{app:60s_continum_spec}), indicating Stark ionization of the Rydberg states under a large, unknown stray electric field. We tried several approaches to mitigate the electric-field issue; however, we observed a broad and unstable Rydberg spectrum only after electrically grounding all peripheral components around the glass cell (see Appendix \ref{app:60s_continum_spec} and \ref{app:grounding}). To better resolve the Rydberg resonance, we switched to the lower $5s44s\,{}^3S_1$ Rydberg state that has been spectroscopically studied \cite{Couturier_Sr-Ryd_2019} and is more robust against electric-field perturbation, as shown in Fig. \ref{fig_setup_scheme}(c). We quickly observed isolated Rydberg resonances in the spectrum; however, the site-resolved spectra were strongly inhomogeneous and temporally unstable across the tweezer array (see Appendix \ref{app:44s_mixed_spectrum}). After several days of repeated diagnostics, the overall spectrum gradually stabilized with resolvable resonances of the $m_J=-1, 0$ manifold, with $m_J$ being the projection of the total electronic angular momentum on the quantization axis, as shown in Fig.~\ref{fig_resolve_ryd_drift}(a). We later attributed the slow reduction in the electric field to weak photo-ionization of the MOT under illumination with a 365 nm UV LED light. 

\begin{figure}[t]
\centering
\includegraphics[width=\columnwidth]{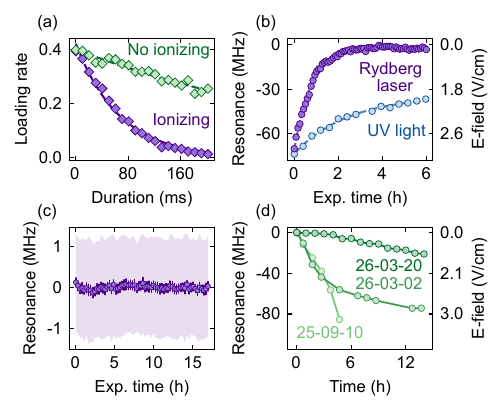}
\caption{\label{fig_neutralize_ryd_stable}
Efficient electric-field reduction with MOT photo-ionization monitored with the $5s44s\,{}^3S_1$ Rydberg state.
(a) Lifetime of the atoms in the MOT after interrupting the pre-cooled atom flux. The tweezer loading rate represents the MOT atom density, indicating rapid depletion by the Rydberg laser. 
(b) Reduction in the stray electric field versus elapsed experimental time, starting from the same field. In each 2.5 s experimental cycle, the Rydberg laser and UV LED light pulses last 200 ms and 1 s, respectively, corresponding to duty cycles of $\sim8\%$ and $\sim40\%$.
(c) Stability in the Rydberg resonance frequency after field elimination, which is better than $\pm0.1$ MHz. The grey background corresponds to the full width at half maximum (FWHM) of the spectrum.
(d) Buildup rate of the residual field after elimination. The rate drops substantially after several months of operation. 
Error bars represent the s.e.m. of the loading rate (a) and the std. of the fitted parameters in the other subplots, and are shown only where they exceed the symbol size.}
\end{figure}

\section{Photo-ionizing elimination of stray electric field} 

With the resolved Rydberg resonance, we can quantitatively compare approaches to eliminating the electric field by monitoring resonance drifts while independently controlling variables such as UV LED illumination, blue-MOT loading, and its repetition rate. From the measurements shown in Fig. \ref{fig_resolve_ryd_drift}(b), we find several important observations. First, when both blue-MOT cooling and UV LED illumination are on (regions I, III, V, VII), the Rydberg resonance drifts toward its zero-field value at rates of -0.17, -0.38, -0.37, and -0.76 V/cm/h, respectively. Second, if only the UV light is switched on while the MOT loading is interrupted (region II, UV desorption only), or the UV light is off while the MOT is running (regions VI, VIII), the resonance drifts rapidly to a large detuning, indicating the buildup of an electric field with rates of 0.37, 0.40, and 0.15 V/cm/h, respectively. Third, when the UV LED is on and the MOT repetition rate is reduced (region IV), the resonance remains roughly unchanged, reflecting a balanced electric field. 

\begin{figure*}[thb]
\centering
\includegraphics[width=0.95\textwidth]{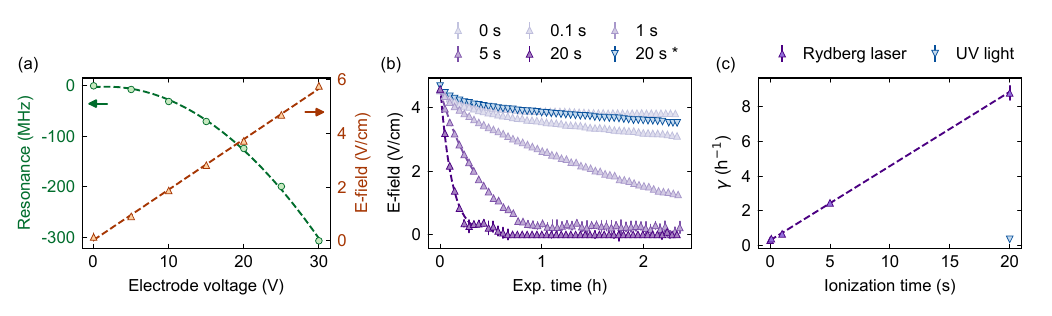}
\caption{  \label{fig_applied_field_benchmark}
Benchmark the MOT photo-ionization method with an externally applied electric field. (a) Calibration of the electric-field strength (orange triangles) versus the absolute applied voltage using the quadratic Stark shift of the $5s44s\,{}^3S_1$ Rydberg resonance frequency (green circles). A linear fit yields a slope of $0.188(3)~\mathrm{(V/cm)/V}$ and an offset of $0.03(6)~\mathrm{V/cm}$. (b) Comparison of the electric-field reduction versus experimental time. We started with a fixed initial electric field of $4.7~\mathrm{V/cm}$, and repeated the Rydberg spectrum scan with a fixed duration of about 3 minutes. Before each spectrum scan, we illuminated the blue MOT with the Rydberg laser for 0, 0.1, 1, 5, or 20 s. Then we recorded the derived electric field versus experimental time. The triangle with an asterisk shows the measurement with UV LED illumination only (20 s) for comparison. (c) Reduction rate in the electric field versus photo-ionization duration. We fit the decay rates in the early parts of the curves in (b) with an exponential function. The rates show a clear linear dependence on photo-ionization duration, with a fitted slope of $0.426(9)~\mathrm{h}^{-1}\mathrm{s}^{-1}$ and an offset of $0.31(2)~\mathrm{h}^{-1}$. In contrast, the UV LED light shows a negligible rate. Error bars represent the std. of the fitted values and are shown only for points where they exceed the symbol size.
}
\end{figure*}

With this observation, we switch to the 317 nm Rydberg laser for comparing the MOT photo-ionization \cite{Witkowski_Sr-ionize_2022}, which has a peak intensity of about 100 W/cm$^2$, whereas that of the UV LED is estimated to be only about 0.5 mW/cm$^{2}$ at the position of the atoms. We first compared the depletion of atoms in the MOT by recording the tweezer loading rate, with the results shown in Fig. \ref{fig_neutralize_ryd_stable}(a). The normal trap lifetime was 420(25) ms and decreased substantially to 66(2) ms under Rydberg laser illumination, indicating efficient photo-ionization of the atoms. We thus estimate that the Rydberg laser produces $\sim 10^3$ ion-electron pairs per ms from the MOT. However, the UV LED light did not noticeably change the MOT lifetime. Figure~\ref{fig_neutralize_ryd_stable}(b) shows the reduction of the stray electric field under periodic photo-ionization. The Rydberg-laser illumination reduces the stray electric field with an exponential time constant of 1.52(2) hours, starting from an initial stray field of about $2.9$ V/cm. Accounting for the duty cycle of the photo-ionization pulse, this $1/e$ decay time corresponds to about 7 minutes of continuous Rydberg laser photo-ionization. The UV LED light has a slower time constant of 4.8(4) hours and saturates at a nonzero value, starting from the same initial field. After field elimination with the Rydberg laser, we repeated the previous sequence and observed a uniform Rydberg transition frequency across the atom array with a stability of about $\pm 0.1~\mathrm{MHz}$ that is much smaller than the width of the spectrum, as shown in Fig. \ref{fig_neutralize_ryd_stable}(c). We further interrupted the photo-ionization procedure and monitored its buildup again, as shown in Fig. \ref{fig_neutralize_ryd_stable}(d). We observed a rapid buildup of the electric field at the beginning, and the rate then decreased substantially after several months.

\section{Benchmark with externally applied electric field}

To better quantify the MOT photo-ionization scheme, we applied equal but opposite voltages to a pair of wire loops wrapped around the objectives to create a controlled DC electric field, as shown in Fig. \ref{fig_setup_scheme}(b). We measured the Stark shift of the $5s44s\,{}^3S_1$ Rydberg resonance frequency at different voltages after clearing the residual electric field, then used it to calculate the field strength with a Stark polarizability of $\alpha_{44s}/h=17.6~\mathrm{MHz/(V/cm)^2}$, with $h$ the Planck constant. The inferred field strength shows a nicely linear dependence on the applied voltage, as shown in Fig.~\ref{fig_applied_field_benchmark}(a), which confirms the zeroing of the background electric field. We then compared the field-reduction effects by photo-ionizing the MOT atoms for variable illumination durations at a fixed electrode voltage of $\pm 25~\mathrm{V}$. To compare the results fairly, we repeated the spectrum scan, fixed the total scan duration to about 3 minutes, and applied a single photo-ionization pulse of variable duration at the beginning of each scan. The recorded reductions in the fitted electric field are shown in Fig.~\ref{fig_applied_field_benchmark}(b). With the Rydberg laser applied to the MOT, the Stark shift quickly approaches zero for sufficiently long repetitions. We clearly observe that the longer the Rydberg laser illuminates the MOT atoms, the faster the remaining electric field decays. For comparison, even 20 s of UV LED light illumination yields only marginal field reduction, similar to the case without any photo-ionization (0 s). In Fig.~\ref{fig_applied_field_benchmark}(c), we plot the exponential reduction rate in the electric field versus the illumination duration in each spectrum scan, and observe a nice linear dependence on its duration, confirming the continuous photo-ionization of MOT-loaded atoms. From the fitted linear coefficient and the illumination duty cycle, the $1/e$ decay time in the electric field corresponds to about 0.8 minutes of continuous Rydberg laser photo-ionization. We note that this measurement corresponds to a background electric field with the slowest buildup rate in Fig. \ref{fig_neutralize_ryd_stable} (d). We indeed observe that the illumination duration required after several months of photo-ionization is nearly one order of magnitude shorter compared to the beginning, as shown in Fig. \ref{fig_neutralize_ryd_stable}(b).

From the chaotic nature of the initial $5s44s\,{}^3S_1$ spectra (see Fig. \ref{fig_44s_chaotic_spectra}), we estimate that the initial stray electric field strength exceeded 10 V/cm by comparing with the Stark map as shown in Fig. \ref{fig_setup_scheme}(c). Such a large stray field may arise from several mechanisms. First, previous studies reveal that alkali adsorbates on dielectric surfaces can create a considerable stray electric field near the surface \cite{Safavi-Naini_chargemodel_2011, Sedlacek_adsorbate_2016}. However, this electric field typically decays as $1/d^3$ with surface distance $d$. Given the 9 mm distance between the atoms and the closest glass surface in our apparatus, this mechanism is unlikely to explain the strong electric field observed. Another possible mechanism is the photoelectric effect induced by the intense 461 nm (2.69 eV) cooling light on the thin adsorbate films accumulated on glass surfaces \cite{Rousseau_laser-charge_1968, Patrick_vapor_2025}. The energy of the blue light is high enough to directly photo-ionize the strontium adsorbates and create patches of surface charges \cite{Gaudart_Sr-film_1977}. The resulting electric field typically decays as $1/d$ and is more likely to be responsible for the large initial electric field. Furthermore, the observed reduction in the accumulation rate of the stray electric field after months of operation, as shown in Fig. \ref{fig_neutralize_ryd_stable}(d), may be explained by the modification in electron affinity on the surfaces \cite{Sedlacek_adsorbate_2016}, and in the work function of the surface photoemission \cite{Gaudart_Sr-film_1977} due to the growth in thickness of the strontium adsorbates. In addition to surface-related effects, we observed significant changes in the electric field after relocating the imaging objective outside the glass cell, pointing to a complex mechanism behind the origin of the stray electric field. Nevertheless, the MOT photo-ionization method presented in this work can efficiently eliminate both the internal surface-related and external-charge-induced electric fields.

\begin{figure}[t]
\centering
\includegraphics[width=0.95\columnwidth]{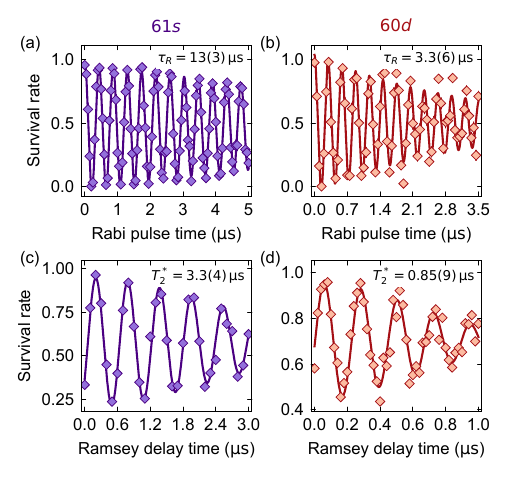}
\caption{\label{fig_rabi_ramsey}
Coherent single-atom excitation of the $5s61s\,{}^3S_1$ and $5s60d\,{}^3D_1$ Rydberg states after field elimination. The upper and lower rows represent resonant Rabi oscillations and delayed Ramsey interference fringes for the two Rydberg states, respectively. The Rabi and Ramsey data are averaged over atoms in a single row with a spacing of $16~\mu\mathrm{m}$. Solid curves represent fits to Rabi oscillations with exponential decay envelopes and Ramsey fringes with Gaussian decay envelopes. The Rabi frequencies for the $5s61s\,^{3}S_1$ and $5s60d\,{}^3D_1$ states are fitted to be $\Omega_{61s}/(2\pi)=2.310(3)~\mathrm{MHz}$ and $\Omega_{60d}/(2\pi)=3.412(9)~\mathrm{MHz}$, with corresponding decay times $\tau_{61s}=13(3)~\mu\mathrm{s}$ and $\tau_{60d}=3.3(6)~\mu\mathrm{s}$, respectively. The Ramsey fits yield detunings $\Delta_{61s}/(2\pi)=1.713(8)~\mathrm{MHz}$ and $\Delta_{60d}/(2\pi)=4.48(4)~\mathrm{MHz}$, together with coherence times $T_{2,61s}^{*}=3.3(4)~\mu\mathrm{s}$ and $T_{2,60d}^{*}=0.85(9)~\mu\mathrm{s}$, respectively. The fast decay of the $5s60d\,{}^3D_1$ state is due to its much higher sensitivity to the electric field (see text). Error bars in the plots denote the s.e.m. of the survival rate and are smaller than the symbol sizes. }  
\end{figure}

\section{Residual electric field after elimination}

After zeroing the electric field, we observed that Rydberg excitation of the $5s61s\,{}^3S_1$ state stabilized to a narrow, uniform profile across the tweezer array (see Appendix~\ref{app:61s_site_spectra}). To evaluate the magnitude of the remaining electric field, we measured the single-atom Rabi oscillation and Ramsey interference pattern between the $5s5p\,{}^3P_0$ clock state and the $5s61s\,{}^3S_1$ and $5s60d\,{}^3D_1\,(m_J=0)$ Rydberg states, with the results shown in Fig.~\ref{fig_rabi_ramsey}. In this measurement, we temporarily switched off the tweezer light during Rydberg-state interrogation. For the $5s61s\,{}^3S_1$ state, we observed high-quality two-level Rabi oscillation, while the Ramsey coherence of this state is primarily limited by thermal motion of the atoms at a temperature of $5\,\mu$K. For the $5s60d\,{}^3D_1$ state, both the Rabi oscillation and Ramsey interference decay much faster than in the $5s61s\,{}^3S_1$ state. Ab initio calculations with PairInteraction give a Stark polarizability of $\alpha_{61s}/h=175~\mathrm{MHz/(V/cm)^2}$. For the $5s60d\,{}^3D_1\,(m_J=0)$ state, the Stark polarizability depends on the relative orientation between the electric and magnetic fields, giving $\alpha_{60d,\parallel}/h=-2174~\mathrm{MHz/(V/cm)^2}$ (parallel) and $\alpha_{60d,\perp}/h=-928~\mathrm{MHz/(V/cm)^2}$ (perpendicular). This larger sensitivity is consistent with the faster Ramsey decay of the $60d$ state. From that, we estimate a conservative upper limit on the standard deviation of the residual electric field as $\delta E=(\pi T_{2,60d}^{*}|\alpha_{60d, \perp}|/h)^{-1/2}=$ 20(1) mV/cm. This stray electric field already meets the requirements of most Rydberg tweezer array experiments.


\section{Summary and Discussion}

In summary, we demonstrated a practical, efficient, and robust method using MOT photo-ionization to reduce detrimental stray electric field drift in a strontium Rydberg tweezer array experiment. Because the origin of the electric field is complex, we infer that in-vacuum photo-ionized charges are drawn to nearby dielectric surfaces under the electric field, reducing it until it’s fully removed. With this method, we eliminated the catastrophic stray electric field of more than 10 V/cm and reduced it to an upper limit of $20(1)~\mathrm{mV/cm}$ in the standard deviation of the remaining DC electric field. This corresponds to a suppression of nearly three orders of magnitude in the electric field. The presented method does not depend on the explicit mechanism behind the origin of the electric field and efficiently addresses both surface-related electric fields and externally static-charge-induced or applied electric fields. Furthermore, this scheme relies only on existing resources, such as laser cooling of atoms and Rydberg lasers or cheap UV LEDs, both readily available in typical ultracold atom experiments. This scheme applies directly to tweezer-array experiments with other atomic species \cite{Anand_dual-species_2024, Mamat_E-field_2024}, circular Rydberg states \cite{Muni_circular_2022, Holzl_circular_2024}, and optical-cavity platforms with Rydberg excitations \cite{Jia_Cavity_2016, Sheng_cavity_2017, Ocola_surface_2024}. A similar scheme may also mitigate stray electric-field issues in polar-molecule or trapped-ion experiments sensitive to uncontrolled static or drifting electric fields \cite{Gempel_mol-electrode_2016, Seesselberg_NaK-electrode_2018, Ziemba_UV-desorp_2025}.


\section*{Acknowledgements}

We thank the USTC strontium optical clock team, especially Yu-Ao Chen, Han-Ning Dai, and De-Quan Kong, for their valuable support. We thank Immanuel Bloch and Lin Li for insightful discussions. This work was supported by the National Natural Science Foundation of China (Grant No. 12274393), the Chinese Academy of Sciences, and the Shanghai Municipal Science and Technology Major Project (Grant Nos. 2019SHZDZX01, 24DP2600300), the Quantum Science and Technology-National Science and Technology Major Project (Grant Nos. 2021ZD0302001), the HFNL Self-Deployed Project (Grant Nos. ZB2024010101, ZB2024010201, and ZB2024010501). Some of the technologies described in this article are included in a patent application (Application No. 202611006517.0), and some of the authors are listed as inventors.

\section*{Data Availability}
The data that support the findings of this article are openly available \cite{data_link}.

\appendix

\setcounter{equation}{0}
\setcounter{figure}{0}
\renewcommand{\theequation}{A\arabic{equation}}
\renewcommand{\thefigure}{A\arabic{figure}}
\setcounter{section}{0}

\section{Steady-state population in MOT}
\label{app:rate-equation-model}

\begin{figure}[!t]
\centering
\includegraphics[width=0.8\linewidth]{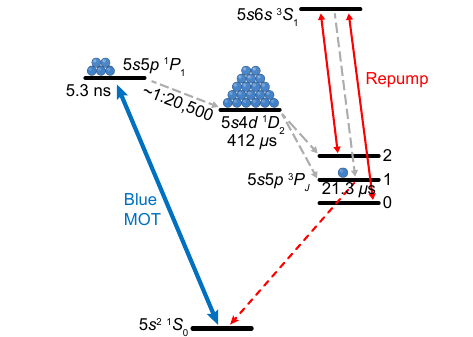}
\caption{\label{fig_steady_state_level} 
Energy-level diagram of $^{88}\mathrm{Sr}$ atoms related to blue-MOT cooling.  The $^1P_1$ decays into the $^1D_2$ state with a branching ratio of $1/20500$ relative to the cooling transition. The lifetimes of the excited $^1P_1$, $^1D_2$ and $^3P_1$ states are $\tau_e$=5.255 ns, $\tau_d=412~\mu\mathrm{s}$ and $\tau_p=21.28~\mu\mathrm{s}$, respectively.
}
\end{figure}

The steady-state population in the excited states during MOT loading is key to the photo-ionizing elimination of stray electric fields. While the excited-state population generally depends on the specific atomic species and laser-cooling scheme, we present the analysis for strontium atoms, with the energy diagram shown in Fig. \ref{fig_steady_state_level}. 

The MOT operates on the broad $5s^2\,{}^1S_0\leftrightarrow5s5p\,{}^1P_1$ transition at 461 nm. During the cooling cycle, a small leakage from the $5s5p\,{}^1P_1$ state populates the long-lived intermediate $5s4d\,{}^1D_2$ state, which then slowly decays into the $5s5p\,{}^3P_J$ manifold, with $J=0,1,2$. We used two intense repumping lights at 679 nm and 707 nm to instantaneously depopulate the atoms from the metastable ${}^3P_0$ and ${}^3P_2$ states into the ${}^3P_1$ state, which slowly decays back to the ground $5s^2\,{}^1S_0$ state. We therefore neglect the populations of the $^3S_1$ and $^3P_{0,2}$ states and keep only four states in the analysis: the ground state $\ket{g}\equiv \ket{{}^1S_0}$, the excited state $\ket{e}\equiv \ket{{}^1P_1}$, the intermediate state $\ket{d}\equiv \ket{{}^1D_2}$, and the intercombination state $\ket{p}\equiv \ket{{}^3P_1}$. For the MOT light with saturation parameter $s=I/I_{\rm sat}$ and effective local detuning $\delta$, the stimulated optical coupling rate between \(\ket{g}\) and \(\ket{e}\) is $W_{ge}=\frac{\Gamma_e}{2}\frac{s}{1+4\delta^2/\Gamma_e^2}$. Here $\Gamma_i=1/\tau_i$ denotes the natural decay rate of state $\ket{i}$ ($i=e,d,p$) and $s=2(\Omega_{ge}/\Gamma_e)^2$, with $\Omega_{ge}$ the Rabi frequency of the cooling light. The rate equations can then be written as,
\begin{equation}
\begin{aligned}
\dot P_e &= W_{ge}(P_g-P_e)-\Gamma_e P_e,\\
\dot P_d &= b\Gamma_e P_e-\Gamma_d P_d,\\
\dot P_p &= \Gamma_d P_d-\Gamma_p P_p,
\end{aligned}
\end{equation}
with $P_i$ the population in state $\ket{i}$, $b$ the branching ratio of the leakage into the $\,{}^1D_2$ state. The normalization of total population requires $P_g+P_e+P_d+P_p=1$. All the derivatives of the steady-state populations are equal to zero and lead to the relation,
\begin{equation}
\begin{aligned}
P_e &= \frac{\Omega^{2}_{ge}}{\Gamma_e^2 + \Omega^{2}_{ge} + 4\delta^2} P_g, \\
b\Gamma_e P_e &= \Gamma_d P_d=\Gamma_p P_p. 
\end{aligned}
\end{equation}

This implies that the excited-state population can be dynamically controlled by tuning the MOT light parameters, and the relative population ratios of the excited states are independent of these parameters. We arrive at the following population ratios,
\begin{equation}
P_e:P_d:P_p=\frac{1}{b\Gamma_e}:\frac{1}{\Gamma_d}:\frac{1}{\Gamma_p} \approx 5:20:1.
\end{equation}

The rate equation shows that the steady-state population mainly accumulates in the intermediate $\,{}^1D_2$ excited state, even though its branching ratio to the $\,{}^1P_1$ excited state is negligible. We note that our Rydberg laser has sufficient single-photon energy to photo-ionize atoms in these three excited states \cite{Haq_Sr-ionize_2006, Haq_Sr-ionize_2007, Witkowski_Sr-ionize_2022}, whereas the ground state cannot be efficiently ionized. In the future, one may identify a more efficient photo-ionization channel for these excited states, given their relative populations.

\section{Stark-ionized $5s61s\,{}^3S_1/5s60d\,{}^3D_1$ Rydberg spectra}
\label{app:60s_continum_spec}

\begin{figure}[htb]
\centering
\includegraphics[width=\columnwidth]{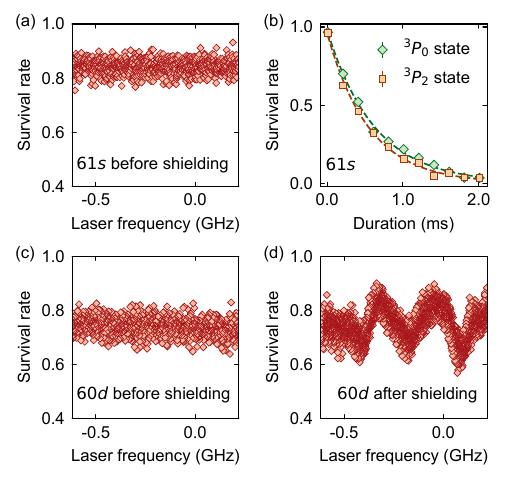}
\caption{\label{fig_ryd_spec_continuum}
Spectroscopy of the Stark-ionized Rydberg spectra of the $5s61s\,{}^3S_1$ and $5s60d\,{}^3D_1$ states under a large stray electric field. (a, c) Remaining fraction of clock-state atoms under a 60 $\mu$s Rydberg laser illumination near the expected resonance frequencies of the states. (b) Lifetime of atoms in the ${}^3P_0$ and ${}^3P_2$ states under far-detuned Rydberg laser illumination. The lifetime of atoms in the clock state is nearly the same as in the ionizable \(^{3}P_2\) state. (d) After grounding nearby components with aluminum foil, we observed unstable loss features in the $5s60d\,{}^3D_1$ spectrum, but could not resolve any isolated resonances. Error bars denote the s.e.m. of the survival rate and are smaller than the symbol sizes. } 
\end{figure}

In our initial attempt to locate the Rydberg transition, we could not resolve any resonances in a broad frequency range near the theoretically predicted frequencies of $5s61s\,{}^3S_1$ and $5s60d\,{}^3D_1$ Rydberg states \cite{madjarov_thesis_2021}, as shown in Fig. \ref{fig_ryd_spec_continuum} (a) and (c). We also monitored the lifetimes of atoms prepared in the $^{3}P_0$ and $^{3}P_2$ states under the illumination of the Rydberg laser at a detuning of $-200$ MHz with respect to the expected resonance frequency of the $5s61s\,{}^3S_1$ state. As shown in Fig. \ref{fig_ryd_spec_continuum} (b), the $^{3}P_2$ state is directly ionizable by the Rydberg laser \cite{Tao_Sr88_2026}, yielding a reasonable lifetime of 0.55(2) ms. However, the observed lifetime of the $^{3}P_0$ state was only 0.63(2) ms, at the same level as the previous state. After examining all relevant laser or magnetic noise, we conjecture that such continuous spectra may arise from Stark ionization of these Rydberg states due to a large, unknown stray electric field. 

\begin{figure}[!t]
\centering
\includegraphics[width=0.95\linewidth]{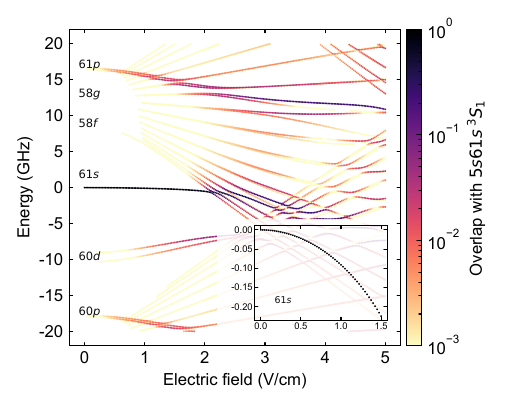}
\caption{\label{fig_61s_stark_map} 
Stark map of the $5s61s\,^{3}S_1$ Rydberg state under an electric field. For the calculation, we include basis states with $56\leq n\leq66$, $0\leq l\leq20$, and $-5\leq m_j\leq5$.  (Inset) Zoom-in of the quadratic Stark shift of the $5s61s\,^{3}S_1$ Rydberg state at a small electric field. }
\end{figure}

Fig. \ref{fig_61s_stark_map} represents the ab-initio calculation of the Stark map of the $5s61s\,{}^3S_1$ and $5s60d\,{}^3D_1$ states by PairInteraction~\cite{Weber_Pair-Interaction_2017}. For highly excited Rydberg states, the energy between adjacent $n$ states as well as the fine-structure interval scales as $n^{-3}$, and the transition dipole moment between neighboring states of opposite parity scales as $n^2$ \cite{Gallagher2009}. Under a weak electric field, the energy of the Rydberg level quadratically shifts due to second-order perturbations from the coupling to neighboring dipole-allowed transitions. As the electric field increases, the Rydberg levels strongly mix with each other and with the ionization continuum. For the $5s61s\,{}^3S_1$ and  $5s60d\,{}^3D_1$ states probed here, an electric field above 3 V/cm essentially Stark-ionizes the Rydberg state. 

\section{Grounding peripheral elements}
\label{app:grounding}

\begin{figure}[ht]
\centering
\includegraphics[width=0.8\linewidth]{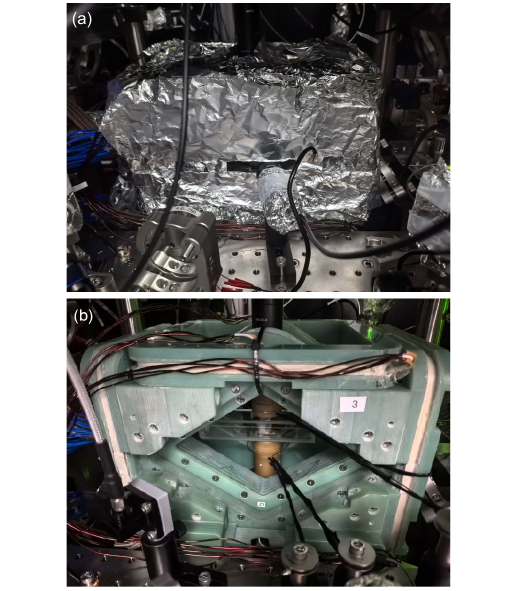}
\caption{\label{fig_shielding} 
(a) Grounded aluminum foil wrapped around the coils and nearby components outside the glass cell to shield external DC electric fields. (b) After photo-ionizing laser-cooled atoms substantially eliminate the stray electric field, it's safe to remove the aluminum foils. We install a pair of loop electrodes to generate an external electric field for benchmarking the neutralization method.}
\end{figure}

To understand the origin of the continuum spectra of Rydberg transitions, as shown in Fig. \ref{fig_ryd_spec_continuum} (a-c), we made a series of attempts before identifying the electric field as the main cause and observing the first evidence. First, we illuminated the glass cell with a 365 nm UV LED at the atom position, as is common in Rydberg tweezer array experiments. Second, we tried blowing ionized air from outside the vacuum chamber toward the glass cell to neutralize static charges that may accumulate there. However, we observed no improvement in the spectra. Third, and most importantly, we wrapped every possible element in thin aluminum foil and properly grounded it (see Fig. ~\ref{fig_shielding}), as the magnetic-field coils were wound around an insulating epoxy mount that could accumulate static charge. 

We also noted that the magnetic-coil power supply was initially floating relative to ground, which may produce a large, oscillating electric field at the power-line frequency. The magnetic coil wiring may also create a voltage gradient across the vacuum chamber, as the current required to drive the clock-state transition reaches 48 A, with a 6 V drop across each coil set. Finite-element simulation shows a tiny electric field up to 0.04 V/cm may appear. To further prevent the field from unknown AC line voltages, we synchronized the experimental sequence to the AC phase. With these efforts, we finally observed broad, unstable structures in the excitation spectrum of the $5s60d\,^{3}D_1$ Rydberg state, as shown in Fig. \ref{fig_ryd_spec_continuum} (d). Although no isolated resonances appear, this confirms that the electric field is the main limitation for the continuum spectra.

\begin{figure}[bt]
\centering
\includegraphics[width=0.9\linewidth]{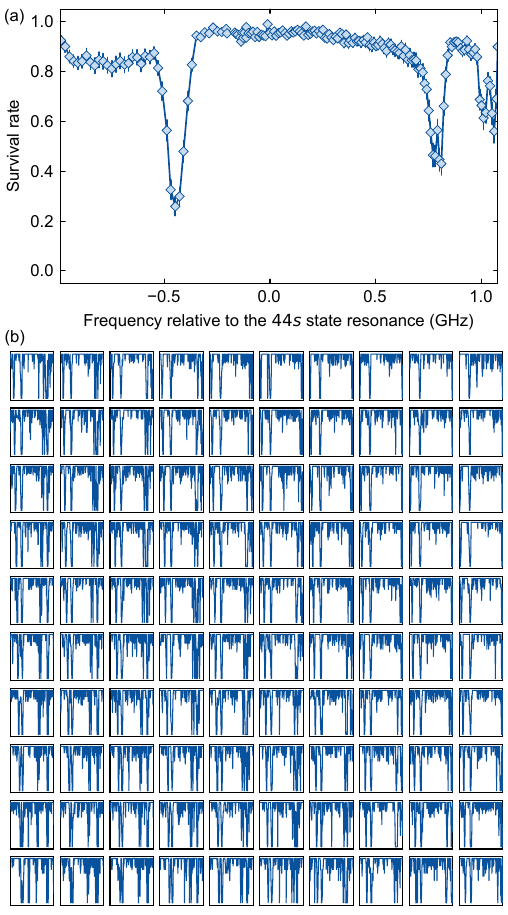}
\caption{\label{fig_44s_chaotic_spectra} 
Mixed and inhomogeneous $5s44s\,^{3}S_1$ Rydberg excitation spectrum disturbed by residual stray electric field after electrical grounding. (a) Averaged spectrum across the atom array. (b) Site-resolved Rydberg spectra corresponding to the averaged spectrum in (a), plotted with the same frequency and survival-rate range. The pronounced site-to-site variation indicates a strongly inhomogeneous and temporally drifting residual electric field across the atom array. Error bars denote the s.e.m. of the survival rate.}
\end{figure}

\section{Mixed $5s44s\,{}^3S_1$ Rydberg spectrum after electrically grounding}
\label{app:44s_mixed_spectrum}

After wrapping the glass cell with grounded aluminum foil and adjusting the laser frequency to the $5s44s\,^{3}S_1$ Rydberg state, the excitation spectrum immediately appeared distinctly from the continuous spectrum of the $5s61s\,^{3}S_1$ state. We stochastically loaded a $10\times10$ tweezer array with $10~\mathrm{\mu m}$ spacing for this measurement. Fig. \ref{fig_44s_chaotic_spectra} (a) shows the array-averaged Rydberg excitation spectrum after electrical grounding. We identified several resonances in the array-averaged survival rate. However, none of the observed resonances coincide with the theoretically expected frequency of the $5s5p\,^{3}P_0 \rightarrow 5s44s\,^{3}S_1$ \cite{Couturier_Sr-Ryd_2019}, indicating the strong Stark shift of the target states and mixing of neighboring Rydberg states by the residual electric field. We can further examine the \textit{in situ} Rydberg spectra at each site of the array, which reflect the local electric fields, averaged over their temporal drift during the measurement, as shown in Fig. \ref{fig_44s_chaotic_spectra} (b). One can clearly observe site-dependent chaotic resonances across the tweezer array that are difficult to interpret, and we attribute them to local gradients in the stray electric field and its temporal fluctuations. We noticed that this gradient pattern shifts across different measurements. As a result, we cannot use such a severely contaminated Rydberg state for further applications.


\section{Uniform Rydberg spectrum after field elimination}
\label{app:61s_site_spectra}

\begin{figure}[ht]
\centering
\includegraphics[width=\linewidth]{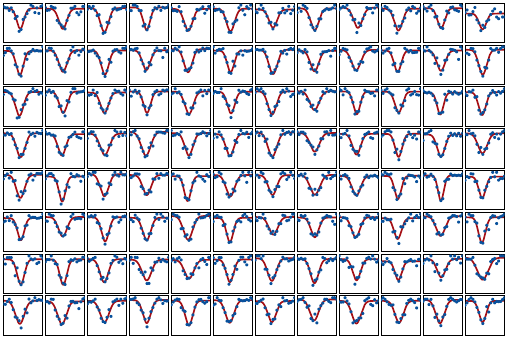}
\caption{\label{fig_61s_uniform_spectra} 
Uniform site-resolved \(5s61s\,{}^3S_1\) Rydberg spectra after field neutralization. The horizontal axis spans a total frequency interval of 4 MHz, and the vertical axis shows the atom survival rate from 0 to 1.}
\end{figure}

After eliminating the stray electric field, we also provide the site-resolved $5s61s\,^{3}S_1$ Rydberg excitation spectra for comparison, as shown in Fig. \ref{fig_61s_uniform_spectra}. We performed this measurement with a stochastically loaded $8\times12$ array with $16\,\mu\text{m}$ spacing. The measured center frequencies exhibit a peak-to-peak variation of 0.27 MHz and a standard deviation of 0.05 MHz, while the average FWHM is 1.0(1) MHz. This measurement proves that even the high-lying $5s61s\,^{3}S_1$ Rydberg state can now be uniformly excited across the atom array. This stability and homogeneity are sufficient for many applications of the highly excited Rydberg states for quantum science.

\bibliography{Ryd_ref}

\end{document}